\documentclass[twocolumn,letter]{jpsj3}
\usepackage{graphicx,bm}

\title{Origin of In-Plane Anisotropy in Optical Conductivity for Antiferromagnetic Metallic Phase of Iron Pnictides}

\author{Koudai {\sc Sugimoto}$^{1,}\thanks{E-mail address: koudai@yukawa.kyoto-u.ac.jp}$, Eiji {\sc Kaneshita}$^{2}$, and Takami {\sc Tohyama}$^{1,3}$ 
}
\inst{$^{1}$Yukawa Institute for Theoretical Physics, Kyoto University, Kyoto 606-8502 \\
$^{2}$Sendai National College of Technology, Sendai 989-3128 \\
$^{3}$JST, Transformative Research-Project on Iron Pnictides (TRIP), Chiyoda, Tokyo 102-0075
}
\recdate{\today}
\abst{We examine the optical conductivity in antiferromagnetic (AFM) iron pnictides by mean-field calculation in a five-band Hubbard model.
The calculated spectra are well consistent with the in-plane anisotropy observed in the measurements, where the optical conductivity along the direction with the AFM alignment of neighboring spins is larger than that along the ferromagnetic (FM) direction in the low-energy region; however, that along the FM direction becomes larger in the higher-energy region.
The difference between the two directions is explained by taking account of orbital characters in both occupied and unoccupied states as well as of the nature of Dirac-type linear dispersions near the Fermi level.}

\kword{magnetically ordered state, iron pnictide, superconductivity, mean-field approximation, optical conductivity, Drude component, anisotropy }

\begin{document}
\sloppy
\maketitle


Iron pnictide superconductors \cite{Kamihara} have recently attracted much attention and have been providing a new direction in the studies of high-temperature superconductivity.
One significant property of iron pnictides is the existence of the antiferromagnetic (AFM) metallic phase in their parent compounds.
This implies a close connection between superconductivity and magnetism in high-temperature superconductors.
In order to elucidate the mechanism of high-temperature superconductivity in iron pnictides, an understanding of the properties of the AFM metallic phase is strongly desired.

Recent resistivity measurements of detwinned samples of the AFM phase~\cite{Tanatar,Chu,Nakajima} in the 122 systems, AFe$_2$As$_2$ (A=Ca, Sr, Ba), have demonstrated anisotropic electric transport in the Fe-As layer:
Below an orthorhombic structural transition temperature, the electric conductivity along the direction with the AFM alignment of neighboring spins (the $x$-direction in our notation below) is larger than that along the ferromagnetic (FM) direction ($y$-direction).
Among the above three compounds, BaFe$_2$As$_2$ has revealed an anisotropy even in a tetragonal phase above the structural transition.
This suggests the presence of a nematic order prior to the structure transition and magnetic ordering, which might be related to the multiorbital nature of iron pnictides.

In addition to the electric resistivity, the optical conductivity of the detwinned BaFe$_2$As$_2$ has shown an anisotropic spectral behavior~\cite{Dusza, Nakajima}:
The optical conductivity along the $x$-direction is larger than that along the $y$-direction below 180~meV while that along the $y$-direction becomes larger above 180~meV.
Together with recent scanning tunneling microscopy measurements~\cite{Chuang}, the study of electronic excitations has revealed a large electronic anisotropy in the AFM metallic phase.
Therefore, it is crucial to clarify the origin of the electronic anisotropy, focusing on the orbital nature of electronic excitations coming from Fe3$d$ bands.

In a theoretical work on optical conductivity by the dynamical mean-field theory (DMFT) combined with first-principles band-structure calculation, Yin \textit{et al.}~\cite{Yin} have reported that the anisotropy in electric transport originates from Pauli's exclusion principle, which blocks the transition to the FM direction.
However, their assertion seems to be too simple to give a full explanation of the anisotropy in the entire energy range.
A very recent fixed-spin-moment band-structure calculation has yielded anisotropic optical conductivity qualitatively consistent with experimental data.~\cite{Sanna}
However, the microscopic origin of the anisotropy has not been fully discussed.
Therefore, no clear explanation has been proposed yet.
To solve this problem, we provide here an interpretation that brings a new viewpoint of the anisotropy based on our calculations.

In this study, we investigate the anisotropy of the optical conductivity by mean-field calculation in a five-band Hubbard model.
Although we use crude approximation as compared with the DMFT work~\cite{Yin}, physical interpretation is easily made to examine the origin of the anisotropy of the optical conductivity. 
Our calculations are consistent with the anisotropic behavior observed experimentally.
Taking account of the symmetry of the orbital character and the presence of Dirac-type linear dispersions, we obtain a physically transparent origin of the anisotropy, which depends on the energy region of the optical conductivity.
We also find that AFM ordering is sufficient to give adequate anisotropy in the optical conductivity, and thus orbital order is not necessary at least below the N\'eel temperature.


Considering an Fe square lattice, we start with a multiband Hubbard Hamiltonian for a $d$-electron system $H=H_0+H_I$.
Here,
\begin{equation}
 H_0 = \sum_{\bm{k},\mu,\nu,\sigma} \Big[ \sum_{\bm{\Delta}} t(\Delta_x, \Delta_y; \mu, \nu) e^{i\bm{k}\cdot{\bm{\Delta}}} + \epsilon_\mu \delta_{\mu, \nu} \Big] c^\dagger_{\bm{k}\mu\sigma} c_{\bm{k}\nu\sigma}
\end{equation}
is the five-band hopping Hamiltonian, where $c^\dagger_{\bm{k}\mu\sigma}$ creates an electron with a wave vector $\bm{k}$ and a spin $\sigma$ at an orbital $\mu$, and $\bm{\Delta}=(\Delta_x,\Delta_y)$. $t(\Delta_x, \Delta_y; \mu, \nu)$ and $\epsilon_\mu$ are the in-plain hopping integrals and on-site energies, respectively, presented by Kuroki \textit{et al.}\cite{Kuroki}.
$H_I$ is the interaction Hamiltonian expressed as\cite{Oles}
\begin{align}
 H_I &= U\sum_{i,\mu} c^\dagger_{i\mu\uparrow} c_{i\mu\uparrow} c^\dagger_{i\mu\downarrow} c_{i\mu\downarrow} \notag \\
      &+ (U-2J) \sum_{i,\mu \not= \nu} c^\dagger_{i\mu\uparrow} c_{i\mu\uparrow} c^\dagger_{i\nu\downarrow} c_{i\nu\downarrow}
	\notag \\
      &+ \frac{U-3J}{2} \sum_{i,\mu \not= \nu,\sigma} c^\dagger_{i\mu\sigma} c_{i\mu\sigma} c^\dagger_{i\nu\sigma} c_{i\nu\sigma}
    \notag \\
      &- J  \sum_{i,\mu \not= \nu} \left( c^\dagger_{i\mu\uparrow} c_{i\mu\downarrow} c^\dagger_{i\nu\downarrow} c_{i\nu\uparrow}
- c^\dagger_{i\mu\uparrow} c_{i\nu\uparrow} c^\dagger_{i\mu\downarrow} c_{i\nu\downarrow} \right),
\end{align}
where $i$ is the Fe-site index, $U$ is the intraorbital Coulomb interaction, $J$ is the Hund coupling, and the pair hopping is set to $J$.
The Fe-Fe bond length is set to unity and the $x$- and $y$-directions are along to the nearest Fe-Fe bonds.
Our calculation uses only two dimensional hopping integrals.
Since interplane hopping integrals are estimated to be small,~\cite{Miyake} this would hardly affect the optical conductivity discussed below.

We self-consistently solve mean-field equations with the order parameter defined by $\langle n_{\bm{Q}\mu\nu\sigma} \rangle = N^{-1}\sum_{\bm{k}} \langle c^\dagger_{\bm{k+Q}\mu\sigma} c_{\bm{k} \nu\sigma} \rangle$ with the ordering vector $\bm{Q}$ and the number of $\bm{k}$ points, $N$, in the first Brillouin zone (BZ) of the paramagnetic phase.
For the observed AFM ordering, we take $\bm{Q}=(\pi,0)$, implying the AFM arrangement of spins along the $x$-direction and the FM arrangement along the $y$-direction.
The very small difference of the bond length between the $x$- and $y$-directions in the orthorhombic phase is not taken into account for simplicity in our calculation~\cite{Kaneshita}.
In addition to $\bm{Q}=(\pi,0)$, the vector $\bm{Q}=(0,0)$ that can represent orbital ordering~\cite{Bascones} is included.
The quasiparticle state $\gamma^\dagger_{\bm{k} \epsilon \sigma} = \sum_{\bm{Q}=(0,0),(\pi,0)}\sum_{\mu} \psi_{\mu\epsilon\sigma}(\bm{k}+\bm{Q}) c^\dagger_{\bm{k}+\bm{Q}\mu\sigma}$ diagonalizes the Hamiltonian with the eigenvalue $E_{\bm{k}\epsilon \sigma}$, where $\epsilon$ is the band index.
The average $\langle \cdots \rangle$ is taken at absolute zero in our calculation.
The computations are performed on a system with $N=500\times 500$.

We set $U=1.2$~eV and $J=0.23$~eV to yield a magnetic moment $M=0.85 \mu_B$ (Bohr magneton) close to the measured moment of the 122 system~\cite{Huang}, where $M = \sum_\mu \langle n_{(\pi,0)\mu\mu\uparrow} - n_{(\pi,0)\mu\mu\downarrow} \rangle \mu_B$.
We note that orbital off-diagonal order parameters result in zero values.
Orbital ordering may be measured using $n_{yz}-n_{zx}$ with $n_\mu=\sum_{\sigma}\langle n_{(0,0)\mu\mu\sigma} \rangle$, whose value is found to be small ($-0.062$).~\cite{Bascones}

Interband contributions to the real part of the optical conductivity are expressed as \cite{Kaneshita}
\begin{align}
 \sigma_{\alpha\beta}(\omega>0) = &\frac{-\pi}{N\omega} \Big(  \frac{e}{\hbar}  \Big)^2 \sum _{\bm{k},\epsilon,\epsilon',\sigma }
 [ f(E_{\bm{k}\epsilon \sigma})- f(E_{\bm{k}\epsilon' \sigma})]
\notag \\
 &\times \zeta^{(\alpha)}_{\bm{k}\epsilon\epsilon'\sigma} [\zeta^{(\beta)}_{\bm{k}\epsilon\epsilon'\sigma}]^* \delta (E_{\bm{k}\epsilon \sigma} - E_{\bm{k}\epsilon' \sigma} - \omega), \label{eq:optcond}
\end{align}
where $e$ is the elementary charge, $f$ is the Fermi distribution function, and $\zeta^{(\alpha)}_{\bm{k}\epsilon\epsilon'\sigma}$ arising from the current operator has the form
\begin{align}
 \zeta_{\bm{k}\epsilon\epsilon'\sigma}^{(\alpha)} = &\sum_{\bm{\Delta},\mu,\nu,\bm{Q}} \Delta^{(\alpha)} t(\Delta_x,\Delta_y;\mu,\nu) e^{-i(\bm{k+Q})\cdot \bm{\Delta}} \notag \\
 & \times \psi_{\mu\epsilon\sigma}(\bm{k+Q})  \psi_{\nu\epsilon'\sigma}^* (\bm{k+Q}),
\end{align}
with $\Delta^{(\alpha)}$ the $\alpha$ component of the vector $\bm{\Delta}$.
The Drude component coming from the intraband transition can be obtained in the limit $\omega \rightarrow 0$ and written as
\begin{align}
 \sigma_{\alpha\beta}(\omega = 0) =&  \frac{1}{8\pi N_F} \Big(  \frac{e^2}{\hbar}  \Big) \sum _{|\bm{k}|=k_\mathrm{F},\epsilon,\sigma} 
 \frac{\zeta^{(\alpha)}_{\bm{k}\epsilon\epsilon\sigma} [\zeta^{(\beta)}_{\bm{k}\epsilon\epsilon\sigma}]^*}{v(\bm{k})},
\end{align}
where $N_F$ is the number of $\bm{k}$ points on the Fermi surface and $v(\bm{k}) = |\partial E_{\bm{k}\epsilon\sigma}/\partial \bm{k}|$.

\begin{figure}
 \begin{center}
 \includegraphics[width=6.0cm]{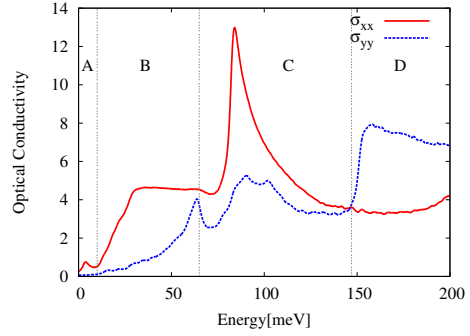}
 \end{center}
 \caption{(Color online)
Interband contribution to the optical conductivity along the $x$-direction $\sigma_{xx}(\omega)$ (red solid line) and $y$-direction $\sigma_{yy}(\omega)$ (blue dashed line).
The Drude component is not included.
The energy scale of the transverse axis is obtained by multiplying the original scale in the mean-field calculation by 1/3 in order to simulate band renormalization due to the correlation effect.
The interband contributions are separated into four regimes labeled as A: range of 0-10~meV, B: 10-65~meV, C: 65-147~meV, and D: 147-200~meV.}
 \label{fig:optcond}
\end{figure}

Figure~\ref{fig:optcond} shows interband contributions to the optical conductivity.
In the figure, we use a renormalized energy scale with a factor $1/3$ to the energy axis, i.e., 0.1~eV in the figure corresponds to 0.3~eV on the original scale of energy.
This factor is taken from comparisons between the dispersion observed by angle-resolved photoemission and the theoretical dispersion determined by first-principles calculation~\cite{Yi,Yoshida}.
The factor corresponds to a band renormalization effect that is included in neither the first-principles calculation nor our mean-field calculation.
The fact that the scaled result in Fig.~\ref{fig:optcond} is consistent with the experimental data (shown below) indicates a crucial role of the self-energy due to the correlation effect.

In Fig.~\ref{fig:optcond}, there are characteristic structures separated into four regimes.
In the range of 0-10~meV (label A), $\sigma_{yy}$ is small (close to zero) and shows no characteristic excitation peak, while $\sigma_{xx}$ shows a peak.
In the range of 10-65~meV (label B), $\sigma_{yy}$ increases more slowly than $\sigma_{xx}$ does.
In addition, there is a plateau in $\sigma_{xx}$.
In the range of 65-147~meV (label C), $\sigma_{xx}$ increases, and a peak appears at 80~meV.
In the range of 147-200~meV (label D), $\sigma_{yy}$ becomes larger than $\sigma_{xx}$ in contrast to that in the lower-energy regimes.

\begin{figure}
 \begin{center}
\includegraphics[width=4.5cm]{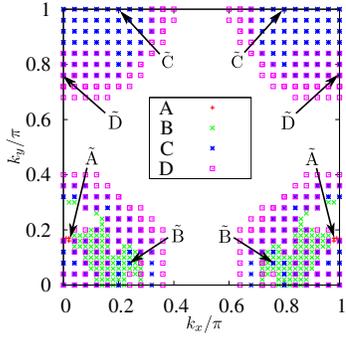}
 \end{center}
 \caption{(Color online) Distribution map of interband transitions contributing to the optical conductivity for each regime (A-D) in the momentum space: A (red $+$), B (green $\times$), C (blue $*$) and D (purple $\Box$). The momentum positions giving rise to dominant contributions for each regime are denoted by \~A to \~D.}
 \label{fig:tr}
\end{figure}

Since the optical conductivity is given by momentum-conserved interband transitions, it is easy to assign momentum regions contributing to regimes of A-D.
Figure~\ref{fig:tr} shows the regions where the energy of direct interband transition across the Fermi level is equal to the energy range of each regime.
Note that the distribution in the momentum space is symmetric with respect to $k_x=\pi/2$ because of the $\bm{Q}=(\pi,0)$ ordering.
Regime A is concentrated in a very narrow region close to the $k_y$ axis.
Regime B is distributed around $\bm{k}=(0.2\pi,0)$ and $(0.8\pi,0)$.
These momenta are close to a crossing point of Dirac-type linear dispersions.~\cite{Ran, Morinari, Richard}
Regime C is assigned to momenta around $(0,0)$ and $(0,\pi)$, and regime D surrounds regime C.
The points denoted by \~A to \~D represent the momenta making a dominant contribution to the largest spectral intensity within each regime, i.e., $\sigma_{xx}$ at $\omega=5$~meV for regime A, $\sigma_{xx}$ at $\omega=25$~meV for regime B, $\sigma_{xx}$ at $\omega=85$~meV for regime C, and $\sigma_{yy}$ at $\omega=160$~meV for regime D.
We find that the positions are located along the highly symmetric line in BZ, except for \~B.
This fact is a key to determining the difference between $\sigma_{xx}(\omega)$ and $\sigma_{yy}(\omega)$, as discussed below.

\begin{figure}
 \begin{center}
 \includegraphics[width=6.0cm]{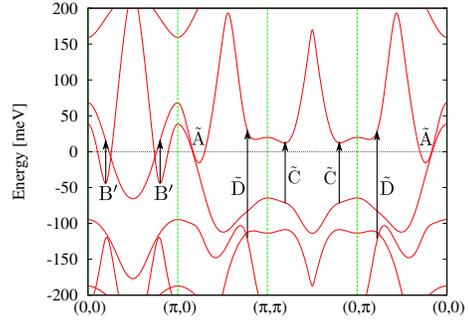}
 \end{center}
 \caption{(Color online) Band dispersions of the $\bm{Q}=(\pi,0)$ AF ordered phases in the five-band Hubbard model. The labels \~A, \~C, and \~D represent the direct excitation process predominantly contributing to the optical conductivity in regimes A, C, and D, respectively. The label B$'$ denotes the position contributing $\sigma_{yy}$ in regime B.}
 \label{fig:eplot2}
\end{figure}

In order to clarify the $\bm{k}$ points contributing to the optical conductivity, we illustrate band dispersions along the symmetric lines in Fig.~\ref{fig:eplot2} together with their positions.
The label B$'$ represents an excitation contributing to a peak of $\sigma_{yy}$ at $\omega=63$~meV in regime B (see Fig.~ \ref{fig:optcond}).

On the symmetric lines in BZ, it is convenient to classify wave functions of the initial and final states in terms of the symmetry of the lines.
The crystal structure of iron pnictides has a $C_2$ rotation about the $x$- and $y$-axes.
When the sign of the orbital wave function changes (remains) under the $C_2$ rotation, we define the parity of the function as $-$ ($+$).
For example, the parity of the $yz$-orbital about the $y$-axis is $-$.
We analyze the wave functions of the initial and final states, and summarize their parity and orbital components in Table~\ref{tab:symmetry}.
The initial and final states are connected by a dipole operator associated with the applied electric field. The parities of the electric field about the $x$-axis rotation are $+$ and $-$ for the $x$- and $y$-polarizations, respectively, and vice versa about the $y$-axis rotation.
Combining the parities of initial and final states with the parity of the electric field, we can judge whether or not the interband transition at \~A to \~D is possible.

\begin{table*}[t]
\caption{Orbitals of initial and final states in each position. The sign in the brackets represents the parity of $C_2$ operation around the rotation axis.}
\begin{tabular}
{cccc}
\hline
 position &  rotation axis & initial & final \\
\hline
\~A & $y$ & $d_{3z^2-r^2}, d_{zx}, d_{x^2-y^2} (+)$ & $d_{yz}, d_{xy} (-) $   \\
B$'$ & $x$ & $d_{zx}, d_{xy} (-)$ & $d_{3z^2-r^2}, d_{yz}, d_{x^2-y^2} (+)$  \\
\~C & $x$ & $d_{zx}, d_{xy} (-)$ & $d_{zx}, d_{xy} (-)$ \\
\~D & $y$ & $d_{yz}, d_{xy} (-)$ & $d_{yz}, d_{xy} (-)$ \\
\hline
\end{tabular}
\label{tab:symmetry}
\end{table*}

\textit{Regime} A.---
$\sigma_{xx}$ mainly comes from \~A on the ($\pi$,0)-($\pi$,$\pi$) line parallel to the $y$-axis in Fig.~\ref{fig:eplot2}.
The orbital of the initial (final) state has the parity $+$ ($-$) about the $C_2$ rotation around the $y$-axis.
The parities of the electric field along the $x$- and $y$-axes are $-$ and $+$, respectively.
Therefore, $\sigma_{yy}$ at \~A vanishes because $+$ (initial) times $+$ (dipole) is not $-$ (final), while $\sigma_{xx}$ at \~A is allowed because $+$ (initial) times $-$ (dipole) is equal to $-$ (final).
This is the main reason why $\sigma_{xx}$ is larger than $\sigma_{yy}$ in regime A.
In this regime, we cannot compare our result with experiments where the Drude component is dominant.
The Drude component is discussed below.

\textit{Regime} B.---
The optical conductivity mainly comes from the momentum space around \~B in Fig.~\ref{fig:tr}.
The spectral behavior in this regime is explained well by the presence of the Dirac dispersions near \~B.
In the vicinity of the Dirac point, the velocity is constant because of linear dispersion, leading to an energy-independent spectral intensity.
This is consistent with the appearance of a plateau in $\sigma_{xx}$.
Note that, since the Dirac point is at $-10$~meV from the Fermi level in the (0,0)-($\pi$,0) direction,~\cite{Morinari} the contribution of this Dirac cone emerges above 10~meV.
In the experimental data, plateaus appear from 45 to 75~meV in both $\sigma_{xx}$ and $\sigma_{yy}$.~\cite{Nakajima}
Our calculations show a plateau in $\sigma_{xx}$, but not in $\sigma_{yy}$.
This also emerges in a previous DMFT work.~\cite{Yin}
In our mean-field calculation, the velocity of the Dirac cone along the $x$-direction is nearly three times as large as that along the $y$-direction.~\cite{Morinari}
Such a difference in the velocity contributes to the difference between $\sigma_{xx}$ and $\sigma_{yy}$.
We should note that the Dirac cones observed in experiments~\cite{Richard} seem to be not very distorted, as calculated.~\cite{Morinari}
Reflecting this, the measured optical conductivity shows less anisotropy in this energy region unlike that in our calculations.
The interband transitions along the (0,0)-($\pi$,0) direction in Fig.~\ref{fig:eplot2} are allowed for $\sigma_{yy}$, as shown in Table~\ref{tab:symmetry}.
As a result, the transition B$'$ appears as a peak at approximately $\omega=65$~meV in $\sigma_{yy}$, as shown in Fig.~ \ref{fig:optcond}.

\textit{Regime} C.---
At \~C in Fig.~\ref{fig:eplot2}, the wave functions of both the initial and final states have the $-$ parity about the $x$-axis rotation, as listed in Table~\ref{tab:symmetry}.
Both states can be connected by the $x$-polarized electric field, but not by the $y$-polarized electric field.
This explains why $\sigma_{xx}$ is larger than $\sigma_{yy}$.
Unlike in the experiments, peak structures emerge at 110~meV in $\sigma_{xx}$ and at 125~meV in $\sigma_{yy}$.~\cite{Nakajima}
These peaks correspond to a sharp peak at approximately 85~meV in $\sigma_{xx}$ and a broad peak at approximately 95~meV in $\sigma_{yy}$ in our calculations.

\textit{Regime} D.---
The main contribution to $\sigma_{yy}$ comes from \~D on the axes parallel to the $y$-direction in Fig.~\ref{fig:eplot2}.
This is similarly explained using the symmetry argument that the initial and final states are connected by the $y$-polarized electric field. 
The opposite anisotropy compared with those of regimes A and C is, thus, caused by different nature of orbitals in the initial and final states.
Our results are consistent with the experimental data showing that $\sigma_{yy}$ is larger than $\sigma_{xx}$ in this regime.~\cite{Nakajima}

The Drude component coming from intraband transition is calculated from the present band structure.
The calculation leads to a larger Drude component along the $x$-direction with the ratio $\sigma_{xx}(\omega=0)/\sigma_{yy}(\omega=0)= 1.3$.
This difference can be understood from the anisotropy of the Fermi velocity.~\cite{Valenzuela} 

In order to extract the effect of orbital ordering on the anisotropy of the optical conductivity, we performed a calculation without the $\bm{Q}=(0,0)$ order parameter (not shown).
We find that the anisotropy remains almost unchanged, and thus conclude that orbital ordering is irrelevant as the origin of the anisotropy but that AFM ordering plays an essential role as the origin. 

In summary, we have investigated the anisotropy of the optical conductivity by mean-field calculation in a five-band Hubbard model.
Our results are consistent with the anisotropic behavior observed experimentally.
The origin of the anisotropy is explained in terms of the orbital characters of the initial and final states contributing to interband transitions. 
In addition to the orbital symmetry, the presence of Dirac-type linear dispersions is crucial for understanding a plateau-like spectral shape in the optical conductivity.
We have found that AFM ordering is enough to give adequate anisotropy in the optical conductivity, and thus an orbital order is not necessary at least below the N\'eel temperature.
The fact that the orbital degree of freedom plays a crucial rule in the AFM metallic phase suggests that the orbital character should be taken into account in further study of superconductivity in pnictides.


We would like to thank M. Nakajima, H. Eisaki, and S. Uchida for stimulating discussions and for providing us their data prior to publication. 
We also thank P. Prelovsek for fruitful discussions.
This work was supported by a Grant-in-Aid for Scientific Research from the MEXT of Japan; the Global COE Program ``The Next Generation of Physics, Spun from University and Emergence"; the Next Generation Supercomputing Project of Nanoscience Program; and Yukawa Institutional Program for Quark-Hadron Science.
The numerical computation in this work was carried out at the Yukawa Institute Computer Facility. E. K. acknowledges a support from Yukawa Memorial Foundation.

\end{document}